\documentclass[aps,twocolumn,a4paper,showpacs,superscriptaddress]{revtex4}
\usepackage{graphicx}
\usepackage{amsmath}
\usepackage{amssymb}
\usepackage{enumerate}
\usepackage{subfigure}
\usepackage{tabularx}
\usepackage[colorlinks=true, pdfstartview=FitV, linkcolor=blue, citecolor=red, urlcolor=black, breaklinks=true]{hyperref}

\newcommand{\be}{\begin{equation}}
\newcommand{\ee}{\end{equation}}
\newcommand{\ben}{\begin{eqnarray}}
\newcommand{\een}{\end{eqnarray}}
\newcommand{\bes}{\begin{subequations}}
\newcommand{\ees}{\end{subequations}}

\newcommand{\bb}{\bibitem}
\newcommand{\sech}{{\rm sech}}
\newcommand{\LL}{{\mathcal L}}

\begin{document}
\title{Generalized scalar field models with the same energy density and linear stability}
\author{L. Losano}\email{losano@fisica.ufpb.br}\affiliation{Departamento de F\'\i sica, Universidade Federal da Para\'\i ba, 58051-970 Jo\~ao Pessoa, PB, Brazil}
\author{M.A. Marques}\email{mam@fisica.ufpb.br}\affiliation{Departamento de F\'\i sica, Universidade Federal da Para\'\i ba, 58051-970 Jo\~ao Pessoa, PB, Brazil}
\author{R. Menezes}\email{rmenezes@dce.ufpb.br}
\affiliation{Departamento de Ci\^encias Exatas, Universidade Federal
da Para\'{\i}ba, 58297-000 Rio Tinto, PB, Brazil}
\affiliation{Departamento de F\'\i sica, Universidade Federal de Campina Grande, 58109-970, Campina Grande, PB, Brazil}
\begin{abstract}
We study how the properties of a Lagrangian density for a single real scalar field in flat spacetime change with inclusion of an overall factor depending only on the field. The focus of the paper is to obtain analytical results. So, we show that even though it is possible to perform a field redefinition to get an equivalent canonical model, it is not always feasible to write the canonical model in terms of elementary functions. Also, we investigate the behavior of the energy density and the linear stability of the solutions. Finally, we show that one can find a class of models that present the same energy density and the same stability potential.
\end{abstract}
\date{\today}
\maketitle

\section{Introduction}
In physics, topological structures appear in several contexts and have been quite investigated over the years \cite{vilenkin,manton,vachaspati}. The most known ones are kinks, vortices and monopoles, which are static solutions of the equations of motion. The simplest structures are kinks, which appear in the presence of scalar fields in $(1,1)$ spacetime dimensions. Usually, they are linearly stable in relativistic field theory. Because of its intrinsec simplicity, kinks can be used, among the many applications, to describe some specific behavior in magnetic materials, superconductors, topological insulators and in other systems in condensed matter \cite{fradkin}. In curved spacetime, kinks may be used to model the fifth dimension in braneworld models with a single dimension of infinite extent \cite{RS,GW,MC,brane,gremm}.

In Ref.~\cite{kinf}, generalized models were introduced in a context of inflation. It was shown that generalized scalar kinetic terms can drive inflationary evolution without the presence of potential terms. Also, they were used to explain why the universe is in an accelerated expansion at a late stage of its evolution \cite{cosm1,cosm2}. Another motivation to study non-canonical models comes from superstring theories. Concerning the tachyon field \cite{asen1,asen2}, the dynamics is modified in a very similar manner to the Born-Infeld concept \cite{bi} of electrodynamics, which introduces nonlinear contributions that works to smooth the divergences of the standard case. In Ref.~\cite{babichev1}, the study of global topological defects was started by considering modifications in the kinetic term of the Lagrangian density. Over the years, defect structures have been vastly studied in generalized models; see Refs.~\cite{kd1,kd2,kd3,kd4,kd5,kd6,kd7,kd8,kd9,bil}.

In this paper, we suggest a class of generalized models and study its properties in flat spacetime, by finding analytical results. In Sec.~\ref{themodel}, we present the model and investigate the fact that, even though a field redefinition to recover the canonical model can be performed, it does not always produce a Lagrangian density that can be explicitly written in terms of elementary functions, which is the scope of the paper. We then go on and search for analytical solutions, as well as their energy densities. To verify the robustness of the model, we also investigate the stability of the kinklike solutions under small fluctuations. We elucidate the possibilities that this new class of models engenders, such as sharing the same energy density and the same stability potential. Finally, in Sec.~\ref{outlook}, we show our conclusions and perspectives.
\section{The model}\label{themodel}
Let us consider the action corresponding to the standard theory for a single scalar field in a $D$-dimensional Minkowski metric, $S=\int d^D \LL$, where
\be\label{eq1}
{\mathcal L} = \frac12 \partial_\mu\chi\partial^\mu\chi -\tilde{V}(\chi)
\ee
and $\tilde{V}(\chi)$ is the potential. In this paper, we introduce a modification in the standard theory by an additional function $f(\phi)$:
\be\label{Lmodel}
{\mathcal L} = f(\phi)\left(\frac12 \partial_\mu\phi\partial^\mu\phi-V(\phi)\right).
\ee
One may wonder if the standard Lagrangian density \eqref{eq1} can be recovered by redefining the field in \eqref{Lmodel}. Indeed, by taking an auxiliary function $h=h(\phi)$ as
\be\label{rebuild}
h_\phi = \sqrt{f(\phi)}\,\,\,\Longrightarrow\,\,\, h(\phi) = \int^\phi{d\tilde\phi \, \sqrt{f(\tilde\phi)}},
\ee
one can show that the change $\chi=h(\phi)$ makes the lagrangian \eqref{Lmodel} become the one in Eq. \eqref{eq1},  where $\tilde{V}(\chi) = V(h^{-1}(\chi))f(h^{-1}(\chi))$.

Of course there are some cases in which the field redefinition is very simple. If we consider, for instance, 
\bes\label{Example1}
\be \label{sinegordon}
V(\phi)=\frac12\cos^2(\phi)
\ee
and 
\be\label{f2}
f(\phi)=\cos^2(\phi), 
\ee
\ees
we obtain $h(\phi)=\sin(\phi)$ which gives $\phi=\arcsin({\chi})$. This will produce $\tilde{V}(\chi) = (1/2) (1-\chi^2)^2$. However, finding an analytic expression for $h(\phi)$ as well as its inverse function is not always possible to be written in terms of elementary functions. This fact is what makes this study interesting.

The scope of this paper is to investigate the model \eqref{Lmodel} deeply. Firstly, we need to understand how the inclusion of the function $f(\phi)$ modifies the dynamics of the scalar field by comparing it to the standard model ($f(\phi)=1$). The equation of motion for the Lagrangian \eqref{Lmodel} can be written as
\be\label{teom}
\Box\phi+V_\phi=-\frac{f_\phi}{f}\left(\frac12 \partial_\mu\phi\partial^\mu\phi+V(\phi)\right)
\ee
and the energy momentum tensor is
\be\label{temgeral}
T_{\mu\nu}=f(\phi)\left(\partial_\mu\phi\partial_\nu\phi - \eta_{\mu\nu}\left(\frac12 \partial_\mu\phi\partial^\mu\phi-V(\phi)\right)\right).
\ee
For static solutions, that is $\phi=\phi(x)$, the equation of motion reduces to
\be\label{secondorder}
\phi^{\prime\prime} = V_\phi-\frac{f_\phi}{f(\phi)}\left(\frac12\phi^{\prime2}-V(\phi)\right).
\ee
By integrating this equation, we get
\be\label{fo}
\frac12\phi^{\prime2}-V(\phi)=\frac{C}{f},
\ee
where $C$ is a integration constant. The above equation allows us to write Eq.~\eqref{secondorder} as 
\be\label{soc}
\phi^{\prime\prime} = V_\phi+C\frac{d}{d\phi}\left(\frac{1}{f}\right).
\ee
This shows that all models with arbitrary $f(\phi)$ have the same solutions only for $C=0$.  Because of that, $\phi^{\prime\prime}=V_\phi$, which is exactly the same of the standard case, that is, $f(\phi)=1$. Coincidently, topological stable solutions must have $C=0$ to have finite energy~\cite{kd1}. Then, the modification present in Eq.~\eqref{Lmodel} does not change the profile of the topological solutions of the standard model. This quantity $C$ is the stress of the solution, $T_{11}$. On the other hand, by using Eq.~\eqref{temgeral}, we obtain the energy density
\be
\rho(x) = T_{00}= f(\phi)\left(\frac12{\phi^\prime}^2 + V(\phi)\right).
\ee
For $C=0$ in Eq.~\eqref{fo}, one can show that the energy density simplifies to
\be\label{rhof}
\rho(x) = 2f(\phi(x)) V(\phi(x))
\ee
From the above equation, we can see that, if we know the solution for the standard model, which happens for $f(\phi)=1$ in Eq.~\eqref{Lmodel}, we also know the energy density for the modified model. We then have a family of models that support the same solutions, but has different energy densities. 

%\section{First-order formalism}
For a general one-dimensional time dependent solution $\phi(x,t)$, we can calculate the energy density and integrate to get the energy
\be
E=\int_{-\infty}^\infty dx f(\phi)\left(\frac12\dot{\phi}^2+ \frac12{\phi^\prime}^2 +V(\phi)\right)
\ee
By using an auxiliary function, $W=W(\phi)$, in the form 
\be\label{defw}
W_\phi=f(\phi)\sqrt{2V(\phi)},
\ee
we can write the energy as
\be
E=\frac12\!\int_{-\infty}^\infty\! dx  f(\phi)\dot{\phi}^2 + \frac12\!\int_{-\infty}^\infty\! dx f(\phi)\! \left(\phi^\prime \mp \frac{W_\phi}{f(\phi)} \right)^2 +E_B,
\ee
with $E_B=|\Delta W| = |W(\phi(\infty))-W(\phi(-\infty))|$. Following the BPS formalism \cite{ps,bogo}, we see each term of the above equation is non-negative. Then, we can ensure that the energy is bounded, $E\geq E_B$. This bound allows us to see that the minimum energy is $E=E_B$ and is achieved for solutions that obey the equations $\dot \phi=0$ (static solutions) and
\be\label{bpseq}
\phi^\prime = \pm \frac{W_\phi}{f(\phi)},
\ee
which, in general, are first-order nonlinear differential equations. We note that $W(\phi)$ may be obtained analytically. The above formalism allows us to find the energies without knowing the solutions. The only information needed to calculate the energy is the minima of the potential.

We take the sine-Gordon potential given by Eq.~\eqref{sinegordon}. The set of topological solutions is
\be\label{solsg}
\phi(x)=\pm\arcsin(\tanh(x))+n\pi,
\ee
connecting the minima $\bar\phi=\pm (2n+1)\pi/2$, where $n\in\mathbb{Z}$.

\begin{figure}[t]
\includegraphics[{width=4.2cm,angle=-00}]{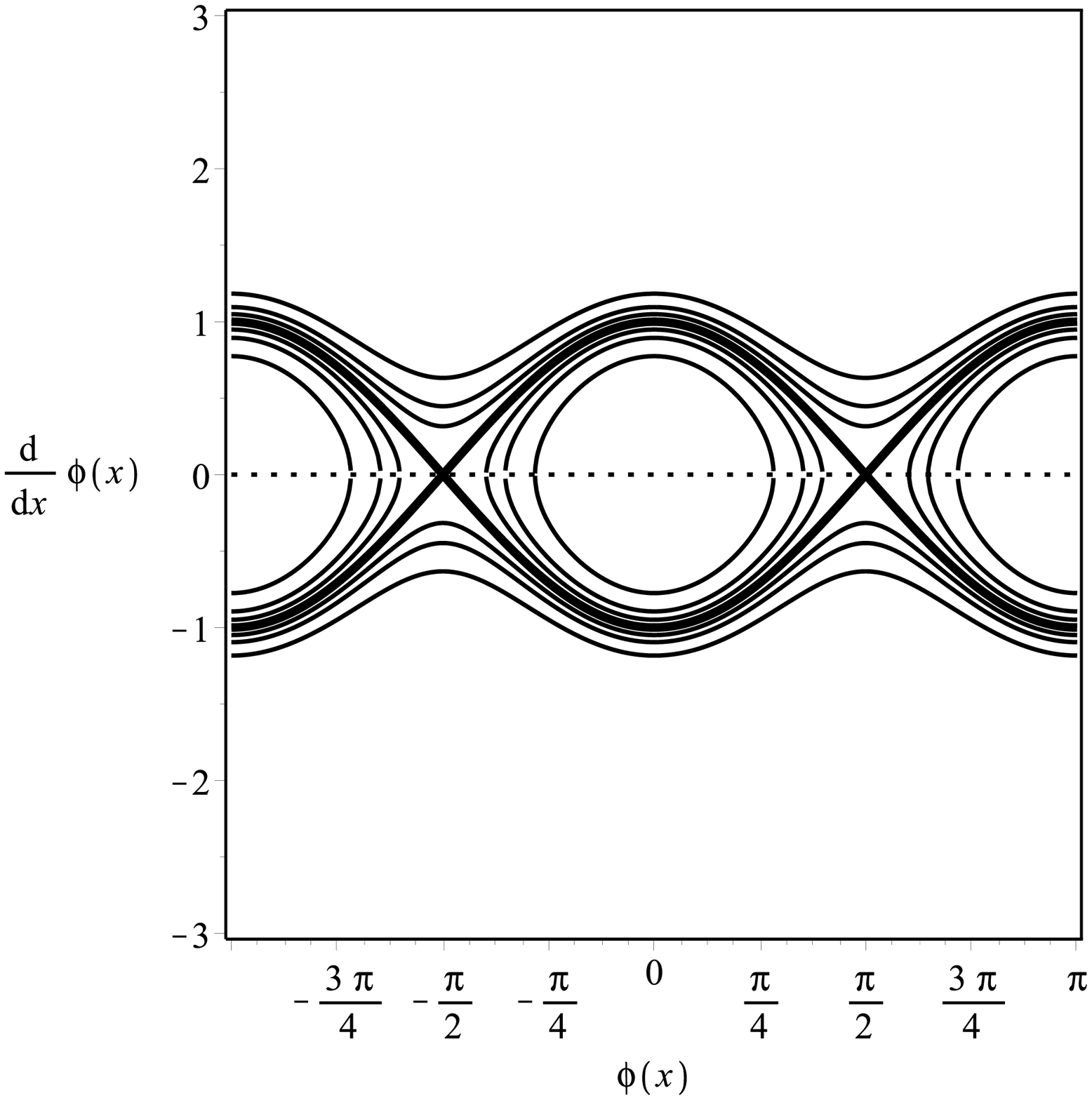}
\includegraphics[{width=4.2cm,angle=-00}]{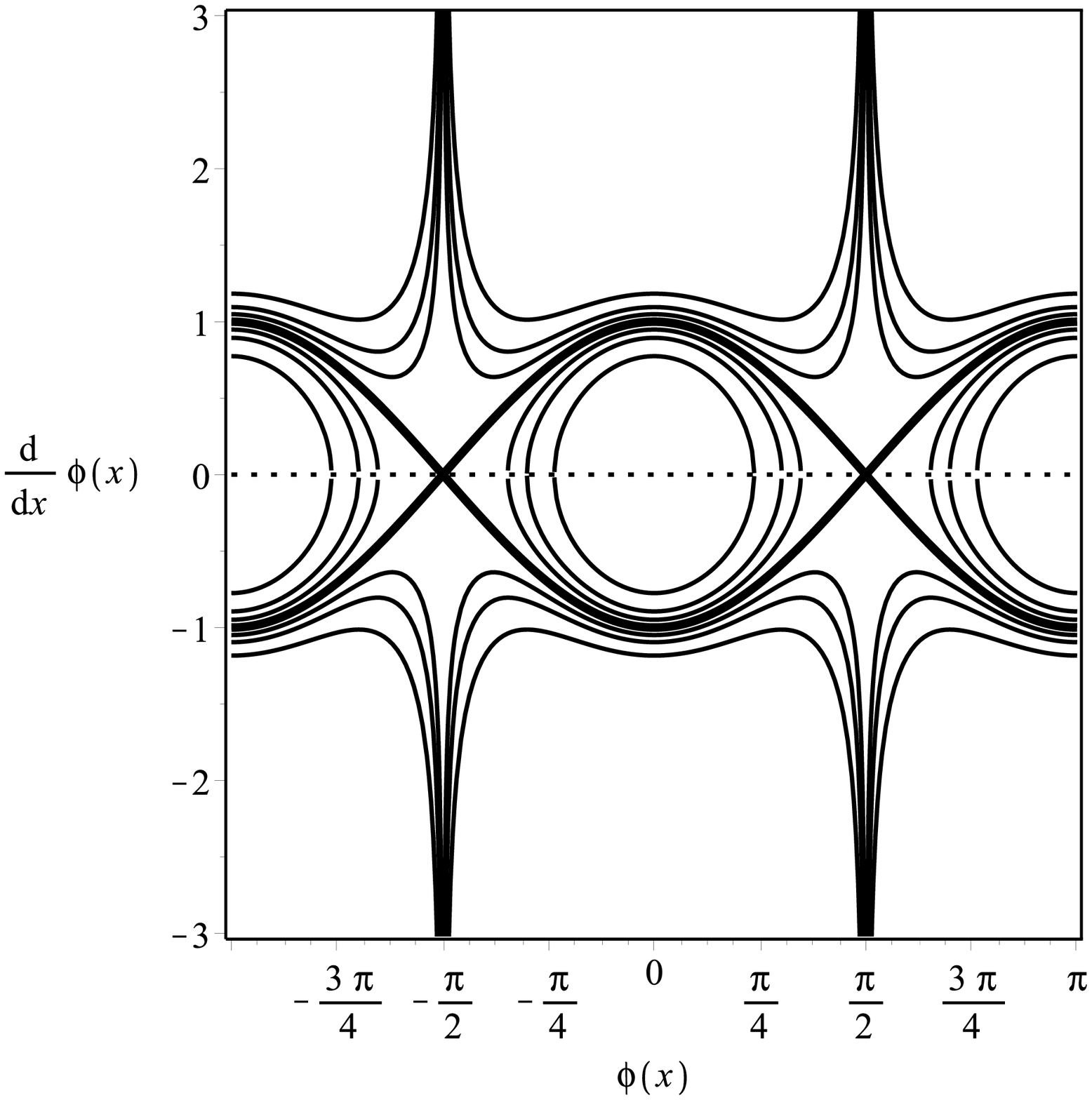}
\includegraphics[{width=4.2cm,angle=-00}]{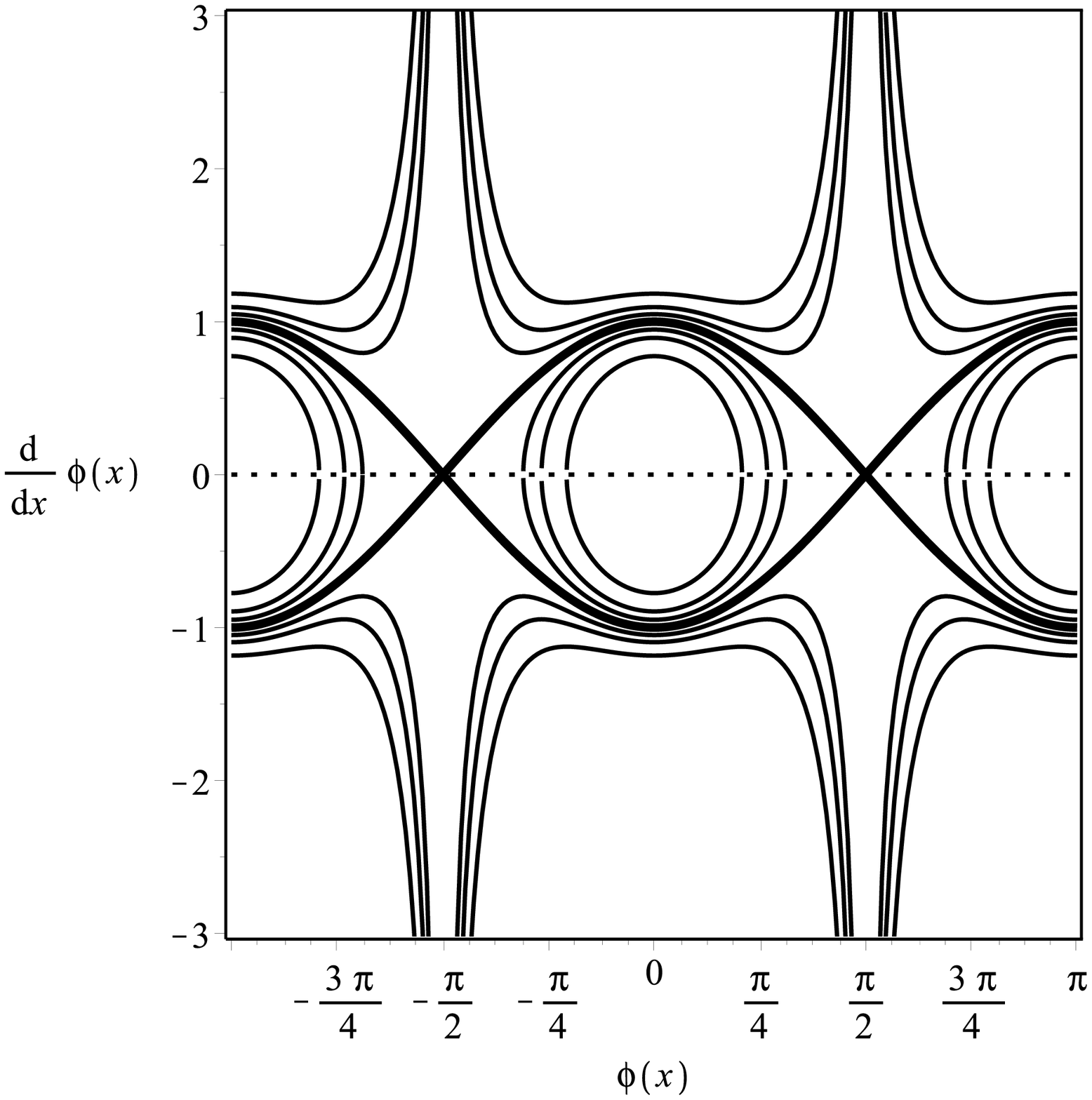}
\includegraphics[{width=4.2cm,angle=-00}]{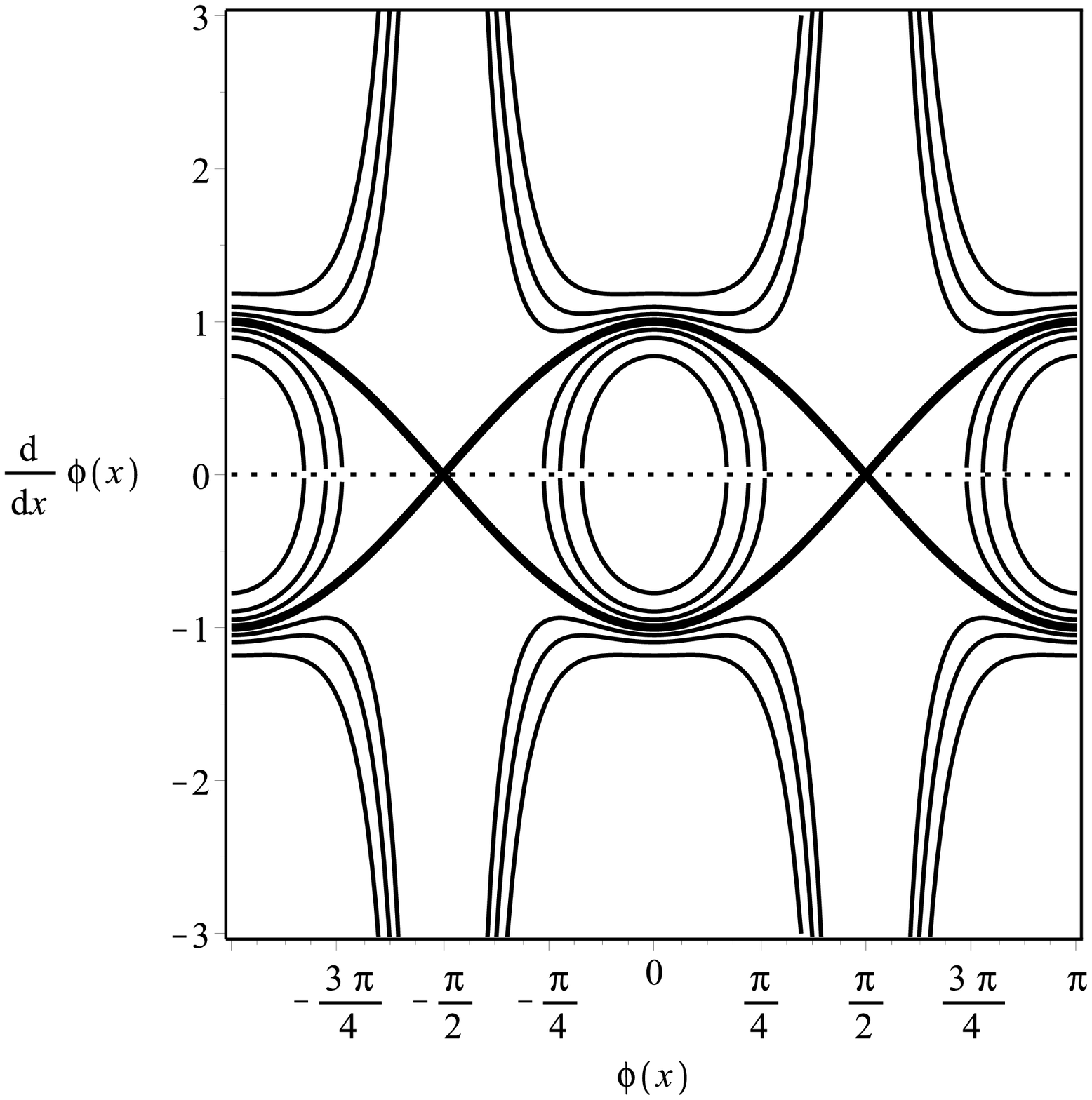}
\caption{The profile of the derivative of the solution as a function of the solution for $f_i(\phi)$, with $i=0,1,2$ and $4$ as in Eqs.~\eqref{feven}. The thickest lines represents the case $C=0$.}\label{cs}
\end{figure}

\begin{figure}[t]
\includegraphics[{width=7cm,angle=-00}]{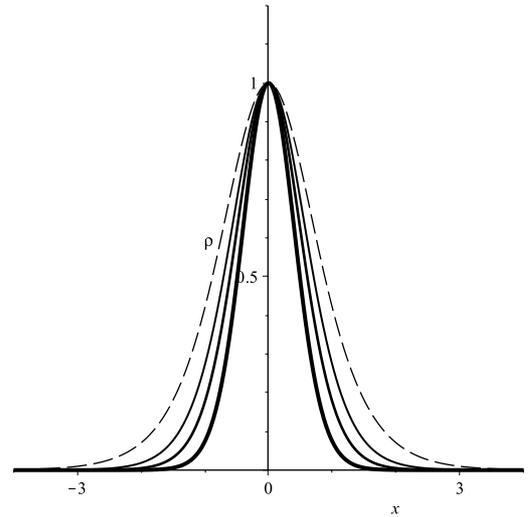}
\caption{The energy densities $\rho_i(x)$, for $i=0,1,2$ and $4$ as in Eqs.~\eqref{rhoeven}. The dashed line stands for the standard case, given by $\rho_0(x)=\sech^2(x)$. The thickness of the lines increases with $i$.}\label{rho}
\end{figure}

In order to analyze which role this new function $f(\phi)$ plays in the game, we consider four different functions:
\bes\label{feven}
\begin{align}
 f_0(\phi)&=1,\\
 f_1(\phi)&=|\cos(\phi)|,\\
 f_2(\phi)&=\cos^2(\phi),\\
 f_4(\phi)&=\cos^4(\phi),
\end{align}
\ees
whose theories admit the solution \eqref{solsg}. Notice that $f_0(\phi)$ is the standard case and $f_2(\phi)$ was considered previously as a trivial case in Eq.~\eqref{f2}. We can use Eq.~\eqref{fo} to plot in Fig.~\ref{cs} the derivative as a function of the field itself for each $f(\phi)$ and several values of the constant $C$.
%%%%%%%%%%%%%%%%%%
%%%%%%%%%%%%%%%%%%%%%%
The graphic shows us that only the curves for $C=0$ are equal, regardless the $f(\phi)$ chosen, as expected. Other values of $C$ shows different behaviors as we change $f(\phi)$.
We use Eq.~\eqref{defw} to find
\bes\label{ws}
\begin{align}
W_0(\phi)&=\sin(\phi) \,\text{sgn}(\cos(\phi)), \\
W_1(\phi)&=\frac{1}{2}\left(\phi + \cos(\phi) \sin(\phi) \right),\\
W_2(\phi)&=\frac13 \sin(\phi) (\cos^2(\phi)+2) \,\text{sgn}(\cos(\phi)), \\
W_4(\phi)&=\frac{1}{15} \sin(\phi) \left(3\cos^4(\phi)+4\cos^2(\phi)+8\right),
\end{align}
\ees 
where $\text{sgn}(\cos(\phi))$ is the signal function of $\cos(\phi)$. Thus $W_0(\pm \pi/2)=\pm 1$, $W_1(\pm \pi/2)=\pm \pi/4$, $W_2(\pm \pi/2)=\pm 2/3$ and $W_4(\pm \pi/2)=\pm 8/15$, which give the energies $E_0=2$, $E_1=\pi/2$, $E_2=4/3$ and $E_4= 16/15$. The energy densities are 
\bes\label{rhoeven}\begin{align}
\rho_0(x)&=\sech^2(x),\\
\rho_1(x)&=\sech^3(x),\\
\rho_2(x)&=\sech^4(x),\\
\rho_4(x)&=\sech^6(x),
\end{align}
\ees
obtained from Eq.~\eqref{rhof}. We plot them in Fig.~\ref{rho}. 
%%%%%%%%
%%%%%%%%

The above examples can be generalized by considering 
\be\label{f2v}
f(\phi)=(2V(\phi))^n,
\ee
where $n$ is a positive real number. In this case, we get the energy density $\rho(x)=T_{00} = 2(2V(\phi(x)))^{n+1}$. Considering the sine-Gordon potential given in Eq.~\eqref{sinegordon}, we get $\rho(x) = 2\, \sech^{2n+2}(x)$. Furthermore, it is possible to find, using Eq.~\eqref{defw}, that
\be
W(\phi) =\,\text{sgn}^{n+1}(\cos(\phi)) \sin(\phi)\, _2F_1\left(\frac12,-n;\frac32;\sin^2(\phi)\right).
\ee
 The above expression leads to the functions obtained in Eq.~\eqref{ws}. Moreover, $W(\pm \pi/2) = \pm \sqrt{\pi}\, \Gamma(n+1)/(2\,\Gamma(n+3/2))$,  which gives the energy $E=\sqrt{\pi}\, \Gamma(n+1)/(\Gamma(n+3/2))$.

\begin{figure}[t]
\includegraphics[{width=4.2cm,angle=-00}]{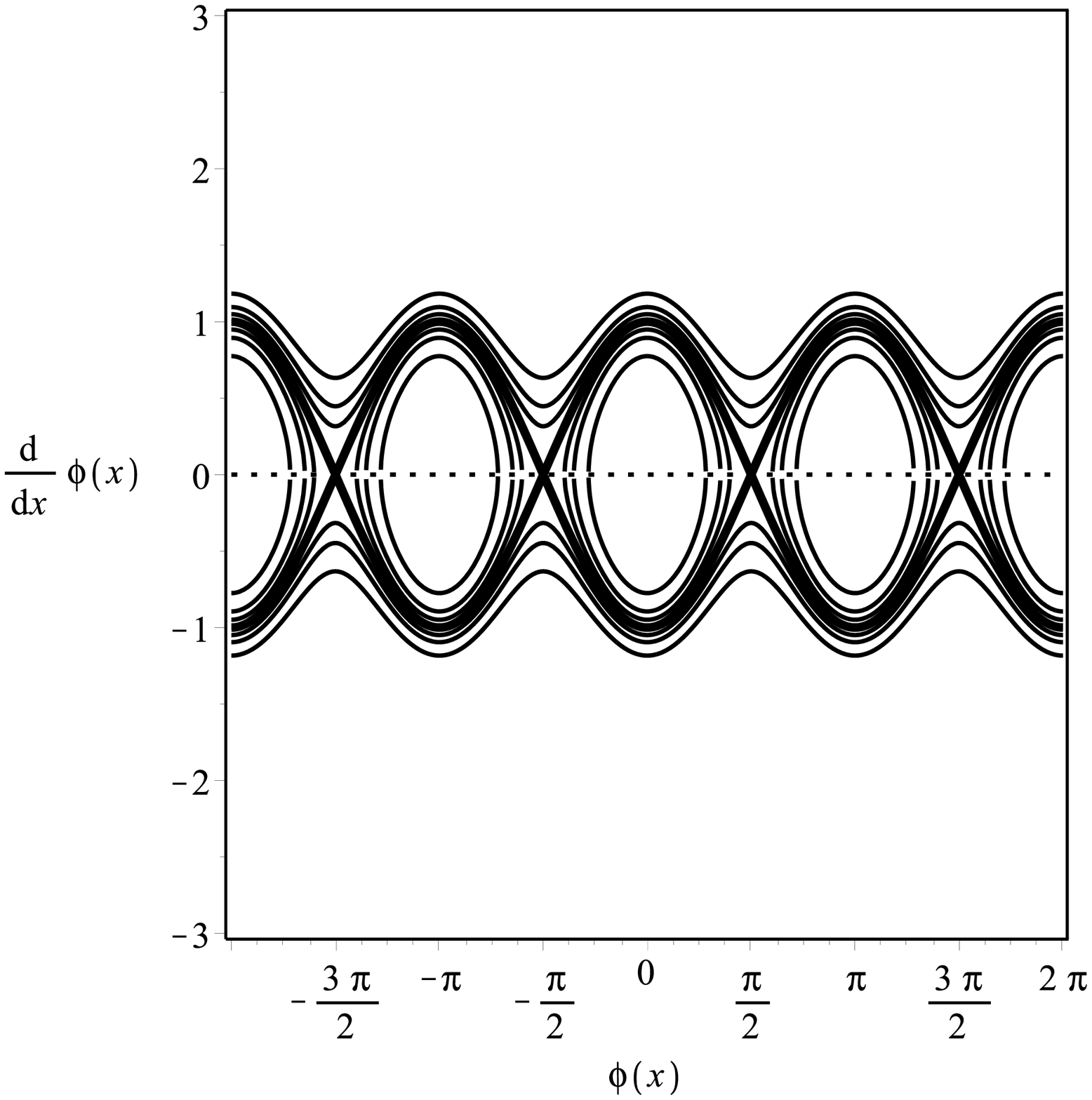}
\includegraphics[{width=4.2cm,angle=-00}]{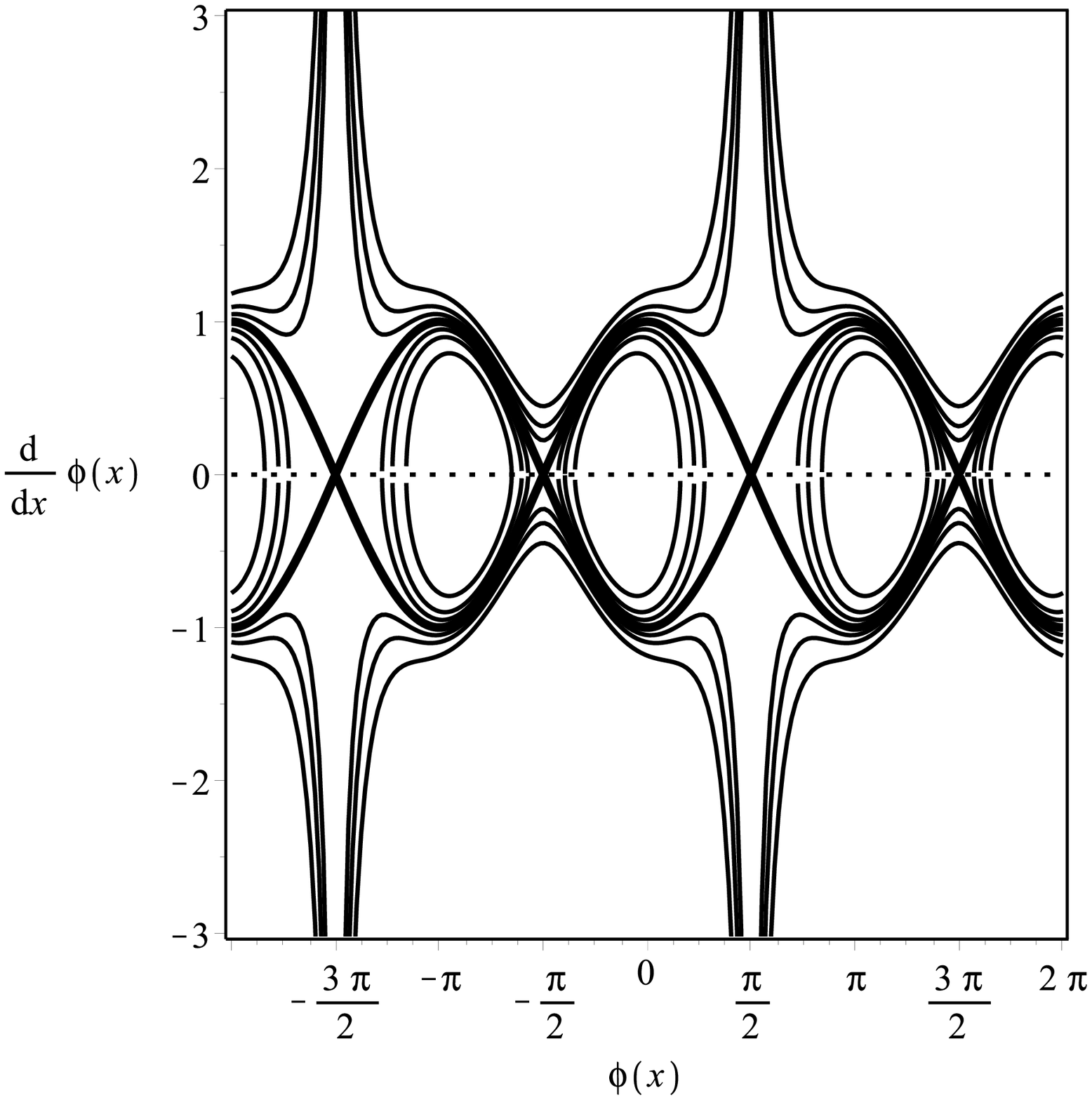}
\includegraphics[{width=4.2cm,angle=-00}]{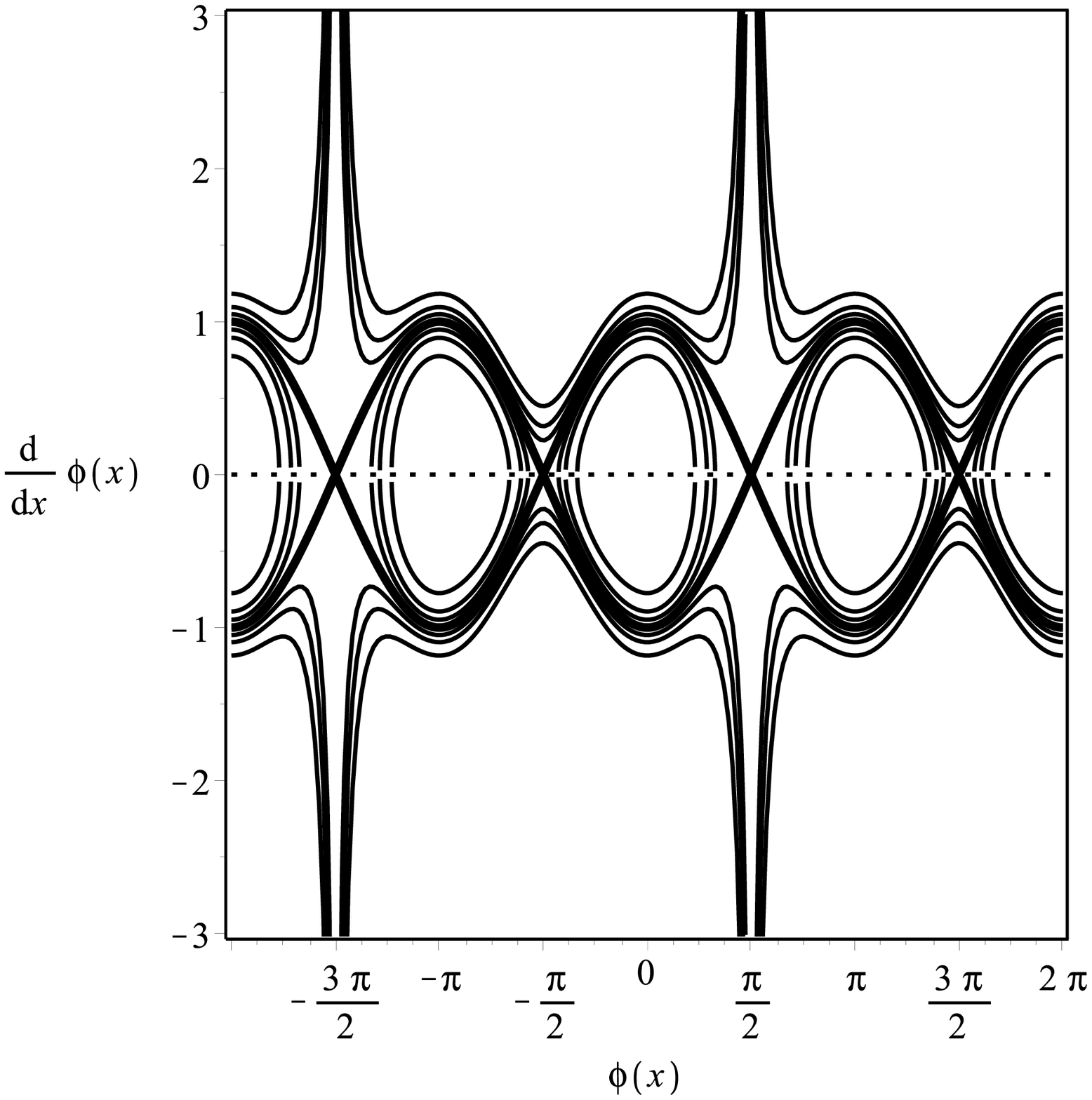}
\includegraphics[{width=4.2cm,angle=-00}]{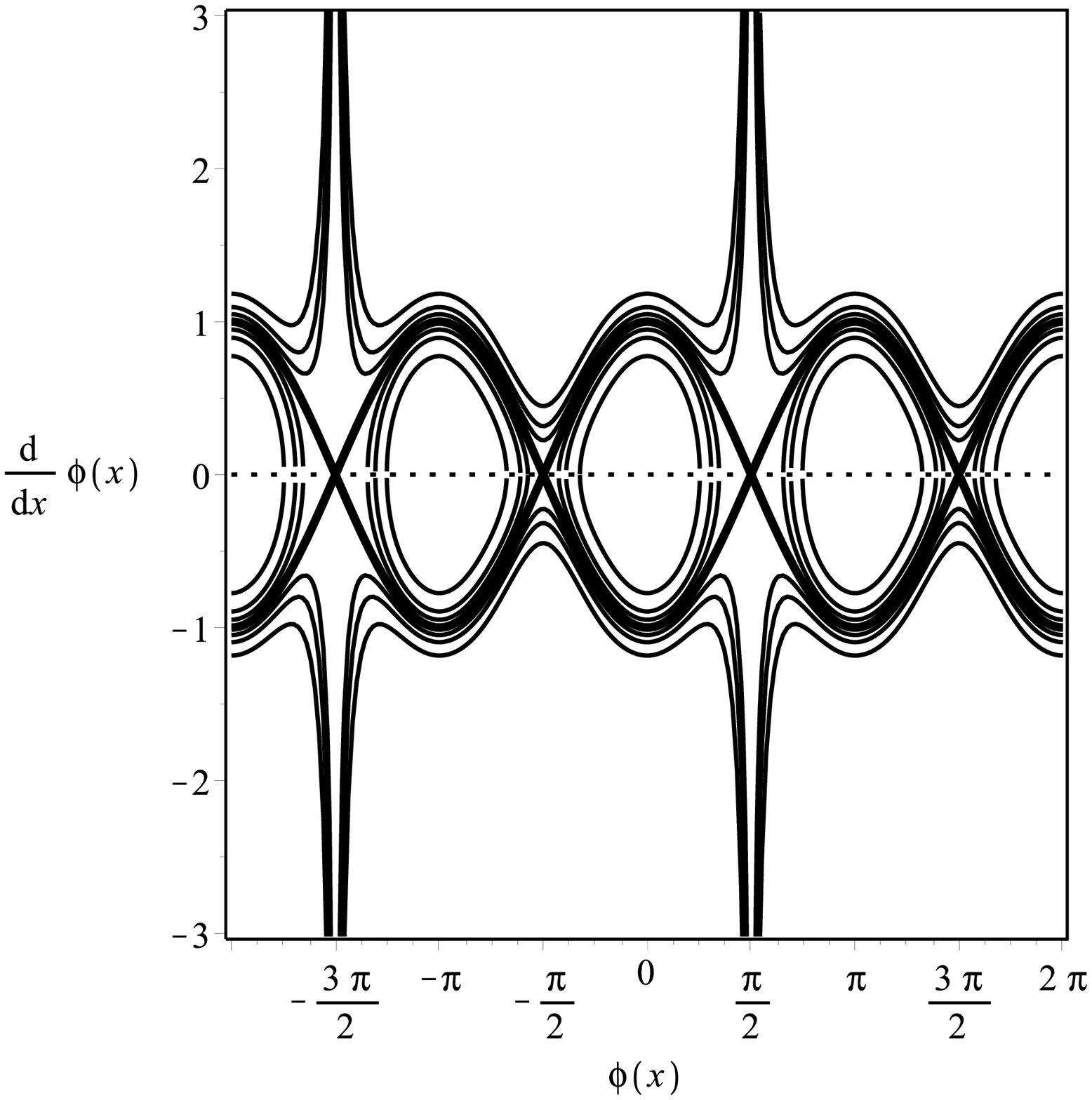}
\caption{The profile of the derivative of the solution as a function of the solution for the standard case $f(\phi)=1$ (top left) and the class of functions \eqref{nenof} for $p=1$ (top right), $p=3$ (bottom left) and $p=5$ (bottom right). The thickest line in each figure represents the case $C=0$.}\label{cs2}
\end{figure}

\begin{figure}[t]
\includegraphics[{width=7cm,angle=-00}]{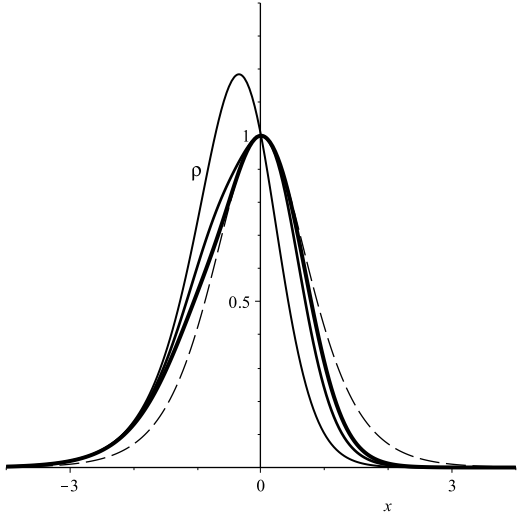}
\caption{The energy densities $\rho(x)$, for the case $f(\phi)=1$, and the others, given by Eq.~\eqref{densneno} for $p=1,3,5$. The dashed line stands for the standard case, given by $\rho_0(x)=\sech^2(x)$. The thickness of the lines increases with $p$.}
\label{rhoneno}
\end{figure}

As another example, we consider, using the potential \eqref{sinegordon}, a different class of functions, which are not even nor odd:
\be\label{nenof}
f(\phi)=1-\sin^{p}(\phi),
\ee
where $p=1,3,5,\dots$. If we take $p=1$ and rebuild its respective canonical field theory as done in Eq.~\eqref{rebuild}, we get $h(\phi)=2\cos(\phi)/\sqrt{1-\sin(\phi)}$. This produces $V(\chi)=\chi^2(8-\chi^2)^2/128$. Again, we can use Eq.~\eqref{fo} to plot in Fig.~\ref{cs2} the derivative as a function of the field itself for each $f(\phi)$ and several values of the constant $C$. Only the curves with $C=0$ are equal, regardless the choice of $p$ in Eq.~\eqref{nenof}. The energy density for this case, from Eq.~\eqref{rhof}, is
\be\label{densneno}
\rho(x)=(1-\tanh^p(x))\sech^2(x),
\ee
which is plotted in Fig.~\ref{rhoneno}. Note that the energy density presents an asymmetric profile. This happens because of the asymmetry of the function \eqref{nenof}.

It is possible to find, by using Eq.~\eqref{defw}, that
\be
W(\phi) =\left( \sin(\phi) - \frac{\sin^{p+1}(\phi)}{p+1}\right) \text{sgn}(\cos(\phi)).
\ee
In the sector $\phi\in[-\pi/2,\pi/2]$, to calculate the energy we must know that $W(\pm\pi/2) = \pm 1 \mp 1/(p+1)$. Then, the energy is $E=2-2/(p+1)$.

\subsection{Linear Stability}

We now study the behavior of the static solution in the presence of small fluctuations. We then take $\phi(x,t) = \phi(x) + \sum_k \eta_k(x) \cos(\omega_k t)$, where $\phi(x)$ is the static solution of Eq.~\eqref{soc}, in the time-dependent equation of motion \eqref{teom} and use Eq.~\eqref{secondorder} to obtain
\ben\label{sturml}
&-&\eta_k^{\prime\prime}- \frac{f_\phi \phi^\prime}{f(\phi)}\eta_k^\prime + \left(V_{\phi\phi} + \frac{f_\phi V_\phi}{f(\phi)} \right)\eta_k \nonumber \\
&+& \left(\frac{f_{\phi}}{f^2} -\frac{f_{\phi\phi}}{f}\right)\left(\frac12{\phi^\prime}^2-V\right)\eta_k = \omega_k^2 \eta_k.
\een
This is a Sturm-Liouville equation. The zero mode is given by $\eta_0(x)=\phi^\prime(x)$. Considering $C=0$, we get
\be
-\eta_k^{\prime\prime} \pm\frac{f_\phi \sqrt{2V}}{f(\phi)}\eta_k^\prime + \left(V_{\phi\phi} + \frac{f_\phi V_\phi}{f(\phi)} \right)\eta_k = \omega_k^2 \eta_k.
\ee
The solution is stable if $\omega_k \in \mathbb{R}$.

We must define a new function, given by $u_k(x) = \sqrt{f(\phi(x))}\, \eta_k(x)$, to transform Eq.~\eqref{sturml} into the Schr\"odinger-like equation:
\be\label{stabeq}
\left(-\frac{d^2}{dx^2} + U(x)\right)u_k(x) = \omega_k^2 u_k(x),
\ee
where $U(x)$ is the stability potential, given by:
%\be
%U(x)=\frac{(\sqrt{{\mathcal L}_X})^{\prime\prime}}{\sqrt{{\mathcal L}_X}} - \frac{1}{{\mathcal L}_X} \left[{\mathcal L}_{\phi\phi} + {({\mathcal L}_{\phi X}\phi^\prime)}^\prime \right].
%\ee
%For our model, we have
\ben
U(x) &=& V_{\phi\phi} -\frac{f_\phi}{f}\left(\frac12\phi^{\prime\prime}-2V_\phi\right) - \frac14\frac{f_\phi^2}{f^2}{\phi^\prime}^2 \nonumber \\
 &&- \frac{f_{\phi\phi}}{f}\left(\frac12{\phi^\prime}^2-V\right) + \frac12 \frac{f_{\phi\phi}}{f}{\phi^\prime}^2.
\een
By using Eqs.~\eqref{secondorder} and \eqref{fo}, we get a stability potential which does not depend explicitly on $C$:
\be\label{stabv}
U(x)= V_{\phi\phi} + \frac32 \frac{f_\phi}{f} V_\phi + \left(\frac{f_{\phi\phi}}{f} - \frac{f_\phi^2}{2f^2} \right) V.
\ee
If we consider $C=0$ in Eqs.~\eqref{secondorder} and \eqref{fo}, we can use \eqref{bpseq} to write the stability potential in terms of the function $W=W(\phi)$:
\begin{align}
U(x) &=\frac{W^2_{\phi\phi}}{f^2} + \left(\frac{W_{\phi\phi\phi}}{f^2}-\frac52 \frac{f_\phi W_{\phi\phi}}{f^3} \right) W_\phi \\
  &\hspace{4mm}+ \left( \frac54 \frac{f^2_\phi}{f^4} - \frac12 \frac{f_{\phi\phi}}{f^3}\right)W_\phi^2.
\end{align}
We can write Eq.~\eqref{stabeq} as
\be
H u_k(x) = \omega_k^2 u_k(x), \quad \text{with} \quad H=S^\dagger S.
\ee
The operator $S$ is given by
\be\label{ops}
S= -\frac{d}{dx} +\frac{W_{\phi\phi}}{f} -\frac12 \frac{f_\phi W_\phi}{f^2}.
\ee
This shows $\omega^2$ is non-negative for a well behaved $W(\phi)$, which implies that the solutions with $C=0$ in Eqs.~\eqref{secondorder} and \eqref{fo} are linearly stable.

The stability potential of the model with the function \eqref{f2v} is:
\be
U(x) = (n+1)V_{\phi\phi} +\frac{n(n+1)}{2} \frac{V_\phi^2}{V}.
\ee
Note that the case $n=0$ reproduces the standard case, $f(\phi)=1$. The operator \eqref{ops} assumes the form
\be
S= -\frac{d}{dx} +\left(n+1\right) \frac{V_\phi}{\sqrt{2V}}.
\ee

Considering the sine-Gordon potential given by Eq.~\eqref{sinegordon} and its solution \eqref{solsg}, we get 
\be
U(x)= \left(n+1\right)^2 - \left(n+2\right)\left(n+1\right)\sech^2(x).
\ee
This is a modified Poschl-Teller potential that admits $n+1$ bound states. Its eigenvalues are given by $\omega_k^2=(2n-k+2)k$, where $0\leq k< n+1$. Furthermore, $U(\pm\infty) = (n+1)^2$. Then,this potential tends to admit more bound states as $n$ increases. This behavior is similar to the one for the compact limit in Ref.~\cite{fktc}.

Now, we turn our attention to the stability of the sine-Gordon potential \eqref{sinegordon} with the function \eqref{nenof}. In this case, the operator \eqref{ops} assumes the form
\be
S=-\frac{d}{dx} +\frac{\left(2\sin^2(\phi)-p\cos^2(\phi)\right)\sin^p(\phi)-2\sin^2(\phi)}{2\sin(\phi)\left(1-\sin^p(\phi) \right)}
\ee
The stability potential \eqref{stabv} becomes
\ben
U(x) &=& 1-2\,S^2(x) +\frac{T^{p-2}(x)\, S^2(x) (3T^2(x)+1)}{1-T^p(x)}\frac{p}{2} \nonumber\\
&& -\frac{T^{p-2}(x)\,S^4(x)(2-T^p(x))}{(1-T^p(x))^2}\frac{p^2}{4} \quad,
\een
where $T(x)=\tanh(x)$ and $S(x)=\sech(x)$. In this case, we have $U(-\infty) = 1$ and $U(\infty) = 4$, vor any $p$. 

Another aspect that we can observe concerning our model is that, for a given potential $V(\phi)$, we can find several functions $f(\phi)$ that lead to the same stability potential. As a special case, we can search which functions $f(\phi)$ lead us to the same stability potential of the standard case, $f(\phi)=1$, which is $U(x)=V_{\phi\phi}$. This can be done by using Eq.~\eqref{stabv} to get the constraint:
\be\label{constraintf}
\frac32\frac{V_\phi}{V} f_\phi  =  \frac{f_\phi^2}{2f} - f_{\phi\phi}.
\ee
It admits the solution
\be\label{ffamily}
f(\phi) = f_0 \,e^{\int\! d\phi\, g(\phi)}\!, \quad g=\left(V^{3/2}\left(B-\frac12\!\int{\!\frac1{V^{3/2}}d\phi}\right)\right)^{-1},
\ee
where  $f_0$ and $B$ are integrating constants. Although, for a given potential $V(\phi)$, the above class of fuctions $f(\phi)$ presents the same solution and the same stability potential of the case $f(\phi)=1$, the energy densities are different.

However, there is another possibility to consider: the function $f(\phi)$ in the Lagrangian density \eqref{Lmodel} gives us the chance to find a set of pairs formed by $f(\phi)$ and $V(\phi)$ that does not change the energy density. This makes Eq.~\eqref{rhof} to become a constraint between $f(\phi(x))$ and $V(\phi(x))$. Nevertheless, as the energy density does not change, the solution changes, since it depends on the potential $V(\phi)$ in Eq.~\eqref{soc} with $C=0$. We can always write the stability potential as a function of the energy density, which is given by Eq.~\eqref{rhof}. Then, Eq.~\eqref{stabv} becomes
\be
U(x)= \frac1{2\rho(x)}\frac{d^2 \rho(x)}{dx^2} - \frac1{4\rho(x)^2}\left(\frac{d\rho(x)}{dx}\right)^2.
\ee
The above potential stability does not change if we choose $f(\phi)$ and $V(\phi)$ to generate the same energy density. This result can lead us to think that there is only one $\rho(x)$ for each $U(x)$. Nonetheless, by considering $\tilde{\rho}(x) =2 f(\phi(x)) V(\phi(x))$, with $f(\phi(x))$ given by Eq.~\eqref{ffamily}, one can show that the stability potential associated to $\tilde{\rho}(x)$ is the same of the standard case, that is $U=V_{\phi\phi}$, because the constraint \eqref{constraintf} is satisfied by $f(\phi(x))$. Therefore, different energy densities can lead to the same stability potentials.

\section{Outlook}\label{outlook}
We have studied how the inclusion of a function of the scalar field in the Lagrangian density modifies the properties of the model. In $(1,1)$ flat spacetime dimensions, the specific form of the Eq.~\eqref{secondorder} allowed us to find models whose stressless solutions are independent of $f(\phi)$. This result cannot be extended for models with more than one field. However, it is possible to perform the same treatment for generalized models with a single scalar field. The generalized action 
\be
S=\int d^2x f(\phi)\mathcal{L}\left(\phi,X\right) \quad \text{with} \quad X=\frac12 \partial_\mu \phi\partial^\mu \phi
\ee
leads to the following equation of motion for static solutions
\be
\left(\mathcal{L}_X \phi^\prime \right)^\prime + \mathcal{L}_\phi = -\frac{f_\phi}{f} \left(\mathcal{L}-2X\mathcal{L}_X \right).
\ee
Again, the right hand of the equation is null for stressless solutions. In this case, the function $f(\phi)$ does not contribute to the equation of motion. Note that, in contrast to the model \eqref{Lmodel}, we cannot always peform a field redefinition to eliminate the function $f(\phi)$. Even though these models share the same field profile, it is clear that the energy density and the linear stability is not same for an specific solution. The study of the time dependence of these solutions is of current interest. One can investigate how the inclusion of the function $f(\phi)$ changes the integrability of known models, as the sine-Gordon model, for instance. Another possibility is to investigate the formation and evolution of domain wall networks. One can also investigate how this function modifies the braneworld scenario with an extra dimension of infinite extent. 

\acknowledgements{We thank Dionisio Bazeia for the discussions that have contributed to this work. We would like to acknowledge the Brazilian agency CNPq for partial financial support. LL thanks support from fundings 307111/2013-0 and 447643/2014-2, MAM thanks support from funding 140735/2015-1 and RM thanks support from fundings 455619/2014-0 and 306826/2015-1.}
%%%%%%%%%%%%%%%%%%%%%%%%%%%%%%%%%

\end{document}